\documentclass{article}
\usepackage{graphicx}
\usepackage{amsmath}
\usepackage{epsfig}
\usepackage{amsfonts}
\usepackage{amssymb}

\newtheorem{theorem}{Theorem}
\newtheorem{acknowledgement}[theorem]{Acknowledgement}

\begin{document}

\title{Square wells, quantum wells and ultra-thin metallic films }
\author{Victor Barsan \\
Department of Theoretical Physics\\
IFIN-HH, Magurele-Bucharest, Romania}
\maketitle

\begin{abstract}
The eigenvalue equations for the energy of bound states of a particle in a
square well are solved, and the exact solutions are obtained, as power
series. Accurate analytical approximate solutions are also given. The
application of these results in the physics of quantum wells are
discussed,especially for ultra-thin metallic films, but also in the case of
resonant cavities, heterojunction lasers, revivals and super-revivals.
\end{abstract}

\section{Introduction}

The present paper illustrates two interesting facts, namely:\ (1) some
important nanophysical systems, for instance quantum wells, can be
understood with elementary quantum mechanics \cite{[Mitin2010]}; and (2)
elementary quantum mechanics offers still unexplored domains, for instance
the transcendental equations of eigenenergy. In fact, in almost all 'exactly
solvable problems' of quantum mechanics, only the Schroedinger equation for
the wave functions has an exact solution, but the transcendental equation
for the eigenvalues can be solved only numerically or graphically. The
attempt of finding analytical - exact or numerical - solutions for these
equations is largely unexplored, even if it might demand sometimes quite
elementary mathematics. So, 'there is plenty of room at the bottom' - to use
Feynman's words in a different context. Inter alia, in this article we shall
undertake such an exercise - to obtain analytical solutions for the
eigenenergy of bound states of a particle in a finite square well. Of
course, the issue of the bound states of a square well has a\ long history,
being discussed in about 40 papers, in the last six decades, to say nothing
about textbooks; for a review, see \cite{[VB-RD]}.

Until the mid '80s, the quantum square well was just a problem of elementary
quantum mechanics, with applications in obtaining the energy levels of
quasi-free electrons moving on long molecules or just below the free surface
of a metal \cite{[Harrison]}, or in the elementary theory of the
Ramsauer-Townsed effect \cite{[Van Wyk2011]}. The odd parity states of the
1D square well are in the same time the $l=0$ solutions of the 3D problem of
a particle moving in a 'radial square well', the so-called 'deuteron
problem', see for instance \cite{[Fluegge1971]}, vol.1, Problem 63, p.162. A
more subtle application refers to ultrathin conducting channels in
field-effect transistors \cite{[Ando1982]}.

How this elementary problem became really important? The impressive progress
of experimental physics in early '80s made possible the fabrication of
heterostructures and quantum wells \cite{[Kolbas1984]}, and the fact that
the general principles of their physics can be easily understood, using the
model of the "particle in a rectangular box" has been immediately exploited.
The basic fact is that, in a very thin film, with planar surfaces, with
mesoscopic or macroscopic longitudinal dimensions, and with nanoscopic
transversal dimensions, the longitudinal momentum (or wave vector) of
electrons is quantized according to cyclic boundary conditions, generating
quasi-continuous values, and the transversal momentum (or wave vector) -
according to rigid boundary conditions, generating quantized values; the
transversal quantum number depends on the film thickness. The transversal
motion of the electrons can be described as a 'quantum square well problem'.
Consequently, the electron spectrum is split in subbands, indexed by the
transversal quantum numbers, with a quasi-continuous energy given by
quasi-continous longitudinal momentum. As the transversal momentum, which
depends on film thickness, cannot be larger than the Fermi momentum, there
is a connection between the film thickness and the Fermi momentum. Due to
the fact that this connection involves discontinous and continous functions,
via the transverse and longitudinal momenta, the physical quantities
describing the system are characterized by functions with discontinous
derivatives, or even with discontinuities. They produce an interesting
physical behavior, namely quantum size effects (QSEs)

In the last two decades, the ultra-thin metallic films where subject of
intense investigation, especially due just to such QSEs, observed in such
systems, consisting of an oscillatory behavior of the film stability \cite%
{[Wu2008]}, of the lattice deformation \cite{[Czosche2003]}, of the work
function \cite{[Kim2010]} etc., as a function of the number of atomic
monolayers. The QSEs, predicted in the pioneering papers of Sandomirskii
\cite{[Sandomirskii1967]} and Schulte\ \cite{[Schulte1976]}, are of
considerable interest, from both practical and theoretical points of view.
Metallic ultra-thin films are especially relevant for ferromagnetic
materials, being responsible for the giant magnetoresistivity observed in
the antiferromagnetic superlattice \cite{[Qiu2002]}. Also, tuning the number
of monolayers means, due to QSEs, tuning the work function, consequently
tuning the chemical properties of the surface \cite{[Kim2010]}.

This quite rich physics can be explained using a quite elementary formalism,
considering, as just mentioned, that the movement of electrons in a
direction transversal to the surface of the metallic film is quantized using
the quantum wells theory. Even the crude approximation of an infinite square
well might be still useful, for instance for the calculation of the lattice
deformation \cite{[Czosche2003]} or of the Fermi energy \cite{[Atkinson2008]}%
. These successes can be easily understood, due to the fact that, if the
number of monolayers is not too small $\left( n\gtrsim 25\right) $, the role
of deep levels is dominant, and the difference between the deep levels of a
finite well and the corresponding levels (i.e. having the same quantum
numbers) of the infinite well is negligible. However, if the number of
monolayers is of the order of unity $\left( n\lesssim 5\right) ,$ it is more
and more inappropriate to approximate the levels of the quantum wells, i.e.
the levels of electrons moving normal to the surface, with the levels of an
infinite well. This is why it is important to obtain the exact values of the
bound states for a well of an arbitrary depth, or reasonably precise
analytic approximations of these quantities.

The structure of this paper is the following. In Section 2, we expose the
elementary problem of the finite square well, discussing carefully the
eigenvalue problem; to find its solution means to obtain a set of functions,
$\xi _{n}$ and $\zeta _{n}$,corresponding to the even and odd parity states.
Section 3 is devoted to the so-called parabolic and cubic approximations for
these functions. In Section 4, we put the transcendental equation for the
energy eigenvalue in a differential form, and obtain an exact series
expansion for $\xi _{n}$ and $\zeta _{n}.$ The 3rd order approximation of
this series - the Barker's approximation \cite{[Barker1991]} - is discussed
in the light of the exact and approximate solutions, as it is already used
in several applications. In Section 5 we discuss the applications of our
results, in three domains: ultra-thin metallic films, revivals and
super-revivals, and electromagnetism. Several concluding remarks are
presented in Section 6.

\section{The quantum square well}

In this section, we shall present the quantum mechanical problem of a
particle in a finite square well. There are in principle two ways of
defining the potential of the well, namely taking the top of the well at $%
E=0,$ or taking the bottom at $E=0.$ In the first case, the advantage is
that the bound states (situated "inside the well") have negative energy -
the typical situation for bound states, in quantum mechanics. In the second
case, the advantage is that, in the limit of a very deep well, the levels
are close to the corresponding levels (having the same quantum number) of
the infinite well. We shall examine in detail the first case, when the
potential is:

\begin{equation}
V\left( x\right) =-U\cdot \theta (\frac{a}{2}-\left\vert x\right\vert )\
\label{1}
\end{equation}%
and $\theta $ defines the Heaviside function. The second case is briefly
discussed later on, see Eqs. (32), (33). The Schroedinger equation for a
particle of mass $m$ moving in the potential (1) has the form:

\begin{equation}
\left[ -\frac{\hbar ^{2}}{2m}\frac{d^{2}}{dx^{2}}+V\left( x\right) -E\right]
\psi \left( x\right) =0\   \label{2}
\end{equation}%
As the potential is invariant to space inversion, $V\left( x\right) =V\left(
-x\right) ,$\ the solutions have well-defined parity. If we introduce the
wave vectors $k,\ k_{0}$ via the relations:

\begin{equation}
E=-\frac{\hbar ^{2}\left( k_{0}^{2}-k^{2}\right) }{2m},\ \ U=\frac{\hbar
^{2}k_{0}^{2}}{2m}\   \label{3}
\end{equation}%
and the dimensionless quantity $P,$ which characterizes both the particle
and the potential$:$

\begin{equation}
P=k_{0}a/2=\sqrt{2mU}\frac{a}{2\hbar }\ ,\ \ p=\frac{1}{P}\
\end{equation}%
Also,

\begin{equation}
\ \frac{U}{P^{2}}=2\frac{\hbar ^{2}}{ma^{2}}\
\end{equation}%
With this notation, the energy of the particle takes the form:

\begin{equation}
E=-U\left[ 1-\left( \frac{ka}{2P}\right) ^{2}\right] \
\end{equation}%
The even solutions of (2) are:

\begin{equation}
\psi _{+}\left( x;k\right) =A_{+}\left( k\right) \left[ \theta (\frac{a}{2}%
-\left\vert x\right\vert )\cdot \cos kx+\right. \
\end{equation}

\begin{equation*}
\left. +\theta (\left\vert x\right\vert -\frac{a}{2})\cdot \cos ka\ \cdot
\exp \left( \sqrt{k_{0}^{2}-k_{n}^{2}}\left( a/2-x\right) \right) \right]
\end{equation*}

\begin{equation}
\psi _{+}\left( -x\right) =\psi _{+}\left( x\right)
\end{equation}%
and the odd ones:

\begin{equation}
\psi _{-}\left( x;k\right) =A_{-}\left( k\right) \left[ \theta (\frac{a}{2}%
-\left\vert x\right\vert )\cdot \sin kx+\right.
\end{equation}

\begin{equation*}
\left. +\theta (\left\vert x\right\vert -\frac{a}{2})\cdot \sin ka\ \cdot
\exp \left( \sqrt{k_{0}^{2}-k_{n}^{2}}\left( a/2-x\right) \right) \right]
\end{equation*}

\begin{equation}
\psi _{-}\left( -x\right) =-\psi _{-}\left( x\right) \
\end{equation}

The explicit form of the normalization constants $A_{\pm }\left( k\right) $
in (7), (9) can be found for instance in \cite{[Fluegge1971]} and will be
not given here.

The energy eigenvalue equations can be conveniently written in terms of the
wave vector $k$. For even states:

\begin{equation}
\ \frac{\cos ^{2}\left( ka/2\right) }{\left( ka/2\right) ^{2}}=\frac{1}{P^{2}%
}=p^{2}\ \
\end{equation}%
and for odd states:\

\begin{equation}
\frac{\sin ^{2}\left( ka/2\right) }{\left( ka/2\right) ^{2}}=\frac{1}{P^{2}}%
\ =p^{2}\
\end{equation}

With $k$ obtained from (11), (12), the eigenenergy is given by (6).

So, to solve the eigenvalue equations (11), (12) means to find the functions
$\zeta \left( p\right) ,\ \xi \left( p\right) $ which satisfy the relations:

\begin{equation}
\frac{\sin \zeta \left( p\right) }{\zeta \left( p\right) }=\pm p,\ \ \frac{%
\cos \xi \left( p\right) }{\xi \left( p\right) }=\pm p\ \
\end{equation}

\bigskip In order to clearly define the alternation of signs in (13), we
shall identify the intervals of monotony of the functions $\sin x/x,\ \cos
x/x.$ Of course, the $x$ varaible used here and hereafter has nothing to do
with the spatial coordinate, enterng in the Schroedinger equation (1) and in
its solutions.The extremum points of the function $\cos x/x$ are given by
the roots $r_{cn}$ of the equation:

\begin{equation}
\tan x=-\frac{1}{x}\
\end{equation}%
where $r_{cn}$\ is the root closest to $\left( n-1\right) \pi :$

\begin{equation}
r_{cn}=\left( n-1\right) \pi -\frac{\left( n-1\right) \pi }{\left(
n-1\right) ^{2}\pi ^{2}-1}\ +\mathcal{O}\left( \frac{1}{n^{2}}\right) \
\end{equation}

The extremum values of the function $\cos x/x$\ are:

\begin{equation}
M_{cn}=\frac{\left( -1\right) ^{n+1}}{\left( n-1\right) \pi }+\mathcal{O}%
\left( \frac{1}{n^{2}}\right) ,\ \ n>1
\end{equation}

It is easy to see that the eigenvalue equations for the even states are:

\begin{equation}
x\in \left( 0,\frac{\pi }{2}\right) :\ \frac{\cos x}{x}=p;\ x\equiv \xi
_{1}\left( p\right) \
\end{equation}

\begin{equation}
x\in \left( r_{c2}\simeq \pi ,\frac{3\pi }{2}\right) :\ \frac{\cos x}{x}%
=-p;\ x\equiv \xi _{2}\left( p\right)
\end{equation}

\begin{equation}
x\in \left( r_{c3}\simeq 2\pi ,\frac{5\pi }{2}\right) :\ \frac{\cos x}{x}%
=p;\ x\equiv \xi _{3}\left( p\right) \
\end{equation}%
and so on.

Similarly, the extremum points of the function $\sin x/x$ are located in the
roots $r_{sn}$\ of the equation:

\begin{equation}
\tan x=x\
\end{equation}%
where $r_{sn}$ is the root closest to $\left( n-\frac{1}{2}\right) \pi .$
The analogous of (15) is:

\begin{equation}
r_{sn}=\left( n-\frac{1}{2}\right) \pi -\frac{1}{\left( n-\frac{1}{2}\right)
\pi }+\ \mathcal{O}\left( \frac{1}{n^{2}}\right) \
\end{equation}

The extremum values of the function $\sin x/x$ are:

\begin{equation}
M_{s1}=1;\ M_{sn}=\frac{\left( -1\right) ^{n-1}}{\left( n-\frac{1}{2}\right)
\pi }+\mathcal{O}\left( \frac{1}{n^{2}}\right) ,\ n>1
\end{equation}

The eigenvalue equations for odd states are:

\begin{equation}
x\in \left( 0,\pi \right) :\ \frac{\sin x}{x}=p;\ x\equiv \zeta _{1}\left(
p\right) \
\end{equation}

\begin{equation}
x\in \left( r_{s,2}\simeq \frac{3}{2}\pi ,2\pi \right) :\ \frac{\sin x}{x}%
=-p;\ x\equiv \zeta _{2}\left( p\right) \
\end{equation}

\begin{equation}
x\in \left( r_{s,3}\simeq \frac{5}{2}\pi ,3\pi \right) :\ \frac{\sin x}{x}%
=p;\ x\equiv \zeta _{3}\left( p\right) \
\end{equation}%
and so on. Each of the Eqs. (17) - (19), (23) - (25) has a unique solution, $%
\xi _{1}\left( p\right) ,\ \xi _{2}\left( p\right) ,\ \xi _{3}\left(
p\right) ,$ respectively $\zeta _{1}\left( p\right) ,\ \ \zeta _{2}\left(
p\right) ,\ \zeta _{3}\left( p\right) ,$ for sufficiently small values of $%
p.\ $On the aforementioned intervals, the functions $\cos x/x,\ \sin x/x$
are monotonous, and for their restriction on each of these intervals, an
inverse function exists. These inverse functions are, in the aforementioned
cases, $\xi _{1}\left( p\right) ,\ \xi _{2}\left( p\right) ,\ \xi _{3}\left(
p\right) ,$ respectively $\zeta _{1}\left( p\right) ,$ $\ \zeta _{2}\left(
p\right) ,\ \zeta _{3}\left( p\right) .$ Their plots can be obtained as
symmetric of the restrictions of the functions $\cos x/x,\ \sin x/x$ on
their intervals of monotony, given by Eqs. (17) - (19), (23) - (25).

Graphically, the functions $\zeta \left( p\right) ,\ \xi \left( p\right) \ $%
correspond to the intersections of the $x-$coordinates of the plots of the
functions $\sin x/x,\ \cos x/x$ with the line $y=\pm p$, on well defined
intervals. The number of solutions depends on the value of $p.$ There is
always at least one solution $\xi \left( p\right) $ for any value of $p$,
but the solutions $\zeta \left( p\right) $ exist only for $p<1.$ In Fig.1,
the plots of the functions $\sin x/x$ (plain),$\ \cos x/x$ (dashed) and $%
y=\pm p,$ with $p=0.1,$ are given. The $x-$coordinate of the intersection
point of $\sin x/x,\ $with $y=p,$ for $0<x<\pi ,$ defines the function $%
\zeta _{1}\left( p\right) ;$ with $y=-p$, for $r_{s,2}\simeq \frac{3}{2}\pi
<x<2\pi \,$, the function $\zeta _{2}\left( p\right) $; with $y=p,$ for $%
r_{s,3}\simeq \frac{5}{2}\pi <x<3\pi ,$ the function $\zeta _{3}\left(
p\right) ,$ etc. Similarly, the $x-$coordinate of the intersection point of $%
\cos x/x,\ $with $y=p,$ for $0<x<\pi /2,$ defines the function $\xi
_{1}\left( p\right) ;$ with $y=-p$, for $r_{c2}\simeq \pi <x<\frac{3\pi }{2}%
\,$, the function $\xi _{2}\left( p\right) $ etc.

As already mentioned, a finite well only supports a fixed number of bound
states. According to several authors (see for instance \cite%
{[AronsteinAJP2000]}), this number is given by:

\begin{equation}
n_{\max }=int\left( \frac{P}{\pi /2}\right) +1\
\end{equation}

In other words, if we define:

\begin{equation}
p\left( N\right) =\frac{2}{\left( N-1\right) \pi }\
\end{equation}%
for $p=p\left( N\right) -0^{+},$ the well supports $N-1$ bound states, and
for $p=p\left( N\right) +0^{+},$ it supports $N$ bound states. This
statement is correct in the approximation

\begin{equation}
r_{cn}=\left( n-1\right) \pi ,\ \ \ r_{sn}=\left( n-\frac{1}{2}\right) \pi \
\end{equation}%
i.e. when we identify the extremum points of the functions $\cos x/x,\ \sin
x/x$ with those of the functions $\cos x,\ \sin x.$ Rigorously speaking, the
situation is slightly more complicated, in the following sense. If $p$
decreases from $\left\vert M_{cn}\right\vert +0^{+}$ to $\left\vert
M_{cn}\right\vert -0^{+},$ a new root, namely $\xi _{n}$ shows up, and this
corresponds to the $\left( 2n-1\right) $-th bound state. Similarly, if $p$
decreases from $\left\vert M_{sn}\right\vert +0^{+}$ to $\left\vert
M_{sn}\right\vert -0^{+},$ a new root, namely $\zeta _{n}$ shows up, and
this corresponds to the $2n$-th bound state.

\bigskip

\begin{figure}[tbph]
\begin{center}
\epsfbox{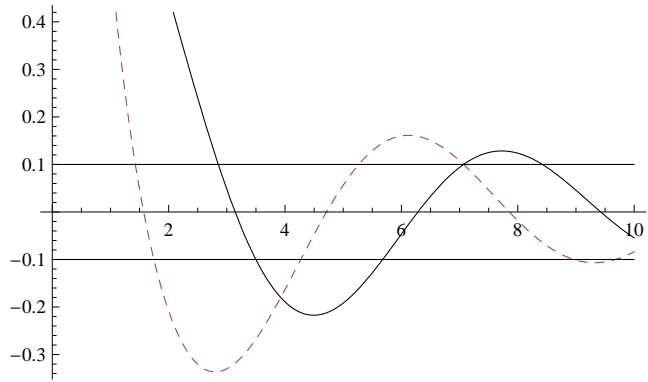}
\end{center}
\caption{The $x-$coordonate of the intersection of the functions $\sin x/x$
(solid) and $\cos x/x$ (dotted) with the lines $y=p$ and $y=-p\,$, according
to Eqs. (23) - (25) and (17) - (19), defines the functions $\zeta _{n}\left(
p\right) $ and $\xi _{n}\left( p\right) ,$ respectively.}
\label{Fig1}
\end{figure}

The $n-$dependence of the coordinates of extremum points, $r_{cn},\ r_{sn}$,
has been obtained by approximating the $\tan $ function in (14), (21) with
the first term of its Taylor series expansion. An alternative way of finding
analytic approximations for the roots $r_{cn},\ r_{sn}$ is to use the
algebraic approximation for $\tan x,$ indicated in \cite{[Alcantara2006]}.
Of course, this is only an example of the usefulness of this algebraic
approximation formula, which can be also used, for instance, to solve the
eigenvalue equation for a particle moving in an infinite rectangular well,
with a $\delta -$function barrier in the middle; indeed, this equation has
the form $\tan z=-z/z_{0}$ \cite{[Alcantara2006]}.

According to Eq. (6), the eigenvalues of the energy are:

\begin{equation}
E_{n}=-U+U\left( \frac{k_{n}a}{2P}\right) ^{2},\ \ n\geqslant 1\
\end{equation}

If we consider a particle which is moving not in the potential $V\left(
x\right) $ given by (1), but in a potential

\begin{equation}
V^{\left( 1\right) }\left( x\right) =V\left( x\right) +U\
\end{equation}%
then the energy levels will be given by:

\begin{equation}
E_{n}^{\left( 1\right) }=U\left( \frac{k_{n}a}{2P}\right) ^{2}=2\frac{\hbar
^{2}}{ma^{2}}\left( \frac{k_{n}a}{2}\right) ^{2},\ \ n\geqslant 1\
\end{equation}

In correspondence to the parity of $n,$ $k_{n}a/2$ corresponds to the
functions $\xi $ or $\zeta ,$ for instance $k_{1}a/2=\xi _{1}\left( p\right)
,\ k_{2}a/2=\zeta _{1}\left( p\right) ,$ etc. In general,

\begin{equation}
E_{2n-1}^{\left( 1\right) }=2\frac{\hbar ^{2}}{ma^{2}}\xi _{n}^{2},\ \
E_{2n}^{\left( 1\right) }=2\frac{\hbar ^{2}}{ma^{2}}\zeta _{n}^{2},\ \
n\geqslant 1\
\end{equation}

Let us remind that the quantity $k_{0}$ defined in (3) cannot be obtained
putting $n=0$ in (29), (31), (32).

So, the eigenfunctions for the even states are:

\begin{equation}
\psi _{2n-1}\left( x\right) =\psi _{+}\left( x;k_{2n-1}\right) =\psi
_{+}\left( x;\frac{2}{a}\xi _{n}\right) ,\ \ n\geqslant 1\
\end{equation}%
and for the odd states:

\begin{equation}
\psi _{2n}\left( x\right) =\psi _{-}\left( x;k_{2n}\right) =\psi _{+}\left(
x;\frac{2}{a}\zeta _{n}\right) ,\ \ n\geqslant 1\
\end{equation}

As it has been already mentioned, the advantage of using the potential (1)
consists in the fact that the energies of the particle "inside the well",
i.e. the energy of the bound states, are negative, which corresponds to the
usual convention adopted in quantum mechanics. However, the form (30) of the
potential has the advantage - also mentioned previously - that its levels
become, in the limit of a very deep well, the levels of the infinite well.
It is indeed easy to see that, for $n\rightarrow \infty ,$ so for very deep
wells, the quantization condition for the wave vector becomes $k_{n}a\simeq
n\pi ,$ so

\begin{equation}
k_{n}\simeq \frac{n\pi }{a}\
\end{equation}%
and the Eq. (31) becomes the equation of the energy levels of an infinite
well:

\begin{equation}
E_{n}^{\left( \infty \right) }=\frac{\pi ^{2}\hbar ^{2}}{2ma^{2}}n^{2}\
\end{equation}

\section{The parabolic and cubic approximations}

The parabolic and cubic approximations consist in approximating the
restrictions of the functions $\cos x/x,\ \sin x/x$ on their monotony
intervals, defined in Section 2, with segments of parabolas, or of cubic
polynomials, having the same roots, extremum points and (in the case of
cubic polynomials) slope in the root, as the exact functions. The detailed
methods for obtaining these approximations have been indicated in \cite%
{[VB-RRP]}, \cite{[VB-RD]}. We shall give here only the results. The
parabolic approximation for the functions $\xi ,\ \zeta $ is:

\begin{equation}
\xi _{n}^{\left( ip\right) }\left( x\right) =r_{cn}+\left( \left( n-\frac{1}{%
2}\right) \pi -r_{cn}\right) \sqrt{1-\frac{x}{M_{cn}}},\ n>1\
\end{equation}

Also,

\begin{equation}
\zeta _{n}^{\left( ip\right) }\left( x\right) \ =r_{sn}+\left( n\pi
-r_{sn}\right) \sqrt{1-\frac{x}{M_{sn}}}\ ,\ n>1
\end{equation}

They are defined on the intervals $\left( M_{\alpha n},0\right) $ for $%
n=even $ and on $\left( 0,M_{\alpha n}\right) ,$ for $n=odd,$ with $\alpha
=c $ for $\xi $ functions and $\alpha =s$ for $\zeta $ functions.

If the extremum points of the functions $\cos x/x,\ \sin x/x$ are
approximated with the extremum points of the functions $\ \cos x,\ \sin x,\ $%
the formulas (37), (38) become:

\begin{equation}
\xi _{n}^{\left( sp\right) }\left( x\right) =\left( n-1\right) \pi +\sqrt{%
\frac{\pi ^{2}}{4}-\frac{x}{4}\left( n-1\right) \pi ^{3}}\ \
\end{equation}

\begin{equation}
\zeta _{n}^{\left( sp\right) }\left( x\right) =\left( n-\frac{1}{2}\right)
\pi +\sqrt{\frac{\pi ^{2}}{4}+\frac{x}{4}\left( n-\frac{1}{2}\right) \pi ^{3}%
}\
\end{equation}%
and their intervals of definition change accordingly. As the formulas (39),
(40) are sometimes referred to as 'simple parabolic approximation', we shall
refer to (37), (38) as 'improved parabolic approximation'. This convention
explains the upper indices in the definition of $\xi _{n}$ and $\zeta _{n}$
functions, Eqs. (37), (38) and (39), (40).

In order to introduce the cubic approximations for the functions $\xi ,\
\zeta ,$ we have to define the quantities:

\begin{equation}
X_{sn,M}=r_{s,n}-n\pi \ ,\ \ \ \ a_{sn,1}=\frac{\left( -1\right) ^{n}}{n\pi }%
\
\end{equation}%
and

\begin{equation}
a_{sn,3}=-\frac{2M_{sn}}{X_{sn,M}^{3}}+\frac{a_{sn,1}}{X_{sn,M}^{2}},\ \ \
a_{sn,2}=\frac{3M_{sn}}{X_{sn,M}^{2}}-\frac{2a_{sn,1}}{X_{sn,M}}\
\end{equation}%
Using the notations:

\begin{equation}
A_{sn,1}=\frac{a_{sn,1}}{a_{sn,3}}-\frac{1}{3}\frac{a_{sn,2}^{2}}{%
a_{sn,3}^{2}}<0;\ \ A_{sn,0}\left( x\right) =\frac{2}{27}\frac{a_{sn,2}^{3}}{%
a_{sn,3}^{3}}-\frac{1}{3}\frac{a_{sn,1}a_{sn,2}}{a_{sn,3}^{2}}-\frac{x}{%
a_{sn,3}}\
\end{equation}%
the cubic approximation for $\zeta _{n}\left( x\right) $ is: \

\begin{equation}
\zeta _{n}^{\left( c\right) }\left( x\right) =n\pi -\frac{a_{sn,2}}{3a_{sn,3}%
}+\frac{2\left\vert A_{sn,1}\right\vert ^{1/2}}{\sqrt{3}}\sin \left( \frac{1%
}{3}\arcsin \left( \frac{\sqrt{27}}{2}\frac{A_{sn,0}\left( x\right) }{%
\left\vert A_{sn,1}\right\vert ^{3/2}}\right) \right) \
\end{equation}

Similarly, defining:

\begin{equation}
X_{cn}=x-\left( n-\frac{1}{2}\right) \pi ,\ \ \ a_{cn,1}=\frac{\left(
-1\right) ^{n}}{\left( n-\frac{1}{2}\right) \pi }
\end{equation}%
and simply replacing the index $s$ by $c$ in (42), (43), we get the cubic
approximation for $\xi _{n}\left( x\right) .$

\begin{equation}
\xi _{n}^{\left( c\right) }\left( x\right) =\left( n-\frac{1}{2}\right) \pi -%
\frac{a_{cn,2}}{3a_{cn,3}}+\frac{2}{\sqrt{3}}\left\vert A_{cn,1}\right\vert
^{1/2}\sin \left( \frac{1}{3}\arcsin \left( \frac{\sqrt{27}}{2}\frac{%
A_{cn,0}\left( x\right) }{\left\vert A_{cn,1}\right\vert ^{3/2}}\right)
\right) \
\end{equation}

The cubic approximations are defined on the same intervals as the
corresponding parabolic approximations.

As $\cos x/x$, for $0<x<\pi /2$ cannot be approximated by a polynomial, the
root $\xi _{1}$ cannot be obtained from the formulas (37) or (46). The
function $\sin x/x$ for $0<x<\pi $ can be approximated by a polynomial, but
cannot be obtained from the general formulas, as the interval of monotony is
not of the form used for the evaluation of $\zeta _{n},$ with $n>1.$
However, the same approach can be used, giving:

\begin{equation}
\zeta _{1}^{\left( p\right) }\left( x\right) =\pi \sqrt{1-x}\
\end{equation}%
for the parabolic approximation (in this case, the simple and the improved
approximation coincide), and

\begin{equation}
\zeta _{1}^{\left( c\right) }\left( x\right) =\frac{\pi \left( 1-x\right)
^{1/2}}{2^{1/2}}\cdot \
\end{equation}

\begin{equation*}
\cdot \frac{1}{-\frac{1}{\sqrt{3}}\sin \left( \frac{1}{3}\arcsin \left(
\frac{3^{3/2}}{2^{5/2}}\left( 1-x\right) ^{1/2}\right) \right) +\cos \left(
\frac{1}{3}\arcsin \left( \frac{3^{3/2}}{2^{5/2}}\left( 1-x\right)
^{1/2}\right) \right) }
\end{equation*}%
for the cubic approximation.

Some caution is needed while switching from $x$ to $p$ dependence in these
formulas. In fact, the parabolic and cubic approximations have been obtained
in a pure geometric context, and the functions $\xi _{n}\left( p\right) ,\
\zeta _{n}\left( p\right) $ in (29), (32) - in a dynamic context, so, in
order to use these approximations in the eigenvalue expressions, we have to
write $\xi _{n}^{\left( c\right) }\left( p\right) \equiv \xi _{n}^{\left(
c\right) }\left( \left\vert x\right\vert \rightarrow p\right) ,\ \zeta
_{n}^{\left( c\right) }\left( p\right) \equiv \zeta _{n}^{\left( c\right)
}\left( \left\vert x\right\vert \rightarrow p\right) .$ The expressions of
the cubic approximations for the functions indexed by $n=2...4$ are given in
the Appendix.

\bigskip

\section{The differential form of the transcendental equations}

We shall expose now a method of solving the eigenvalue equations, discovered
and re-discovered several times, which generates an exact solution, in the
form of a series expansion. As we shall see, the method works for a large
class of transcendental equations.

Let us take the derivatives in both sides of the equations (17)-(19) and
(23)-(25), putting:

\begin{equation}
X_{2n}\left( p\right) =\zeta _{n}\left( p\right) ,X_{2n-1}\left( p\right)
=\xi _{n}\left( p\right) ,\ n=1,2,...,\ \ \
\end{equation}%
replacing $p$ by $x$ and relaxing the restriction $p>0.$ The equations for
the energy eigenvalues can be written in a unitary form:

\begin{equation}
\frac{dX_{n}\left( x\right) }{dx}=-\frac{X_{n}\left( x\right) }{\sqrt{%
1-x^{2}X_{n}\left( x\right) ^{2}}+x}\
\end{equation}%
with the initial condition:

\begin{equation}
X_{n}(0)=\frac{n\pi }{2}
\end{equation}

It is remarkable that both the even, and the odd solutions, satisfy
identical differential equations.The formula (50) allows us to obtain the
derivatives of any order of $X_{n}\left( x\right) $\ in an arbitrary point $%
x_{0}$ and, consequently, to construct the Taylor series for this function,
with arbitrary precision. Choosing $x_{0}=0$, and putting:

\begin{equation}
\frac{X_{n}\left( x\right) }{X_{n}\left( 0\right) }=\frac{X_{n}\left(
x\right) }{n\pi /2}=x_{n}\left( x\right) =\sum_{m=0}^{\infty }q_{m}\left(
\frac{n\pi }{2}\right) x^{m}\
\end{equation}

we get the following exact series expansion for $x_{n}\left( x\right) :$

\begin{equation}
x_{n}\left( x\right) =\sum_{m=0}^{\infty }q_{m}\left( \frac{n\pi }{2}\right)
x^{m}\
\end{equation}

The first 6 functions $q_{m}$ are the following polynomials in the variable

\begin{equation}
\frac{n\pi }{2}\equiv b_{n},
\end{equation}%
namely

\begin{equation}
q_{0}\left( b_{n}\right) =1,\ \ q_{1}\left( b_{n}\right) =-1,\ \ q_{2}\left(
b_{n}\right) =1\
\end{equation}

\begin{equation}
q_{3}\left( b_{n}\right) =-\left( 1+\frac{b_{n}^{2}}{6}\right) ,\ \
q_{4}\left( b_{n}\right) =\left( 1+\frac{2b_{n}^{2}}{3}\right) \
\end{equation}

\begin{equation}
q_{5}\left( b_{n}\right) =-\left( 1+\frac{5}{3}b_{n}^{2}+\frac{3}{2^{3}\cdot
5}b_{n}^{4}\right) ,\ \ q_{6}\left( b_{n}\right) =\left( 1+\frac{2\cdot 5}{3}%
b_{n}^{2}+\frac{2^{3}}{3\cdot 5}b_{n}^{4}\right) \
\end{equation}

The polynomials $q_{2n-1}\left( b\right) ,\ q_{2n}\left( b\right) $ have the
same order, $2n-1.$ We calculated quite a lot of such polynomials (see the
Appendix) with the expectation that this exercise could help to guess a
general formula for its coefficients. However, the presence of some large
prime numbers in the products which enter in the coefficient of polynomials $%
q_{m}$, specially for odd indices, is somewhat discouraging for the attempt
of finding general expressions for this quantities.

The limit $p\rightarrow 0$ corresponds to the infinite well. It is easy to
see, using (35) and (53), that:

\begin{equation}
E_{n}^{\left( 1\right) }\left( p\right) =E_{n}^{\left( \infty \right) }\cdot
x_{n}^{2}\left( x\right) \
\end{equation}%
with

\begin{equation}
E_{n}^{\left( \infty \right) }=2\frac{\hbar ^{2}}{ma^{2}}X_{n}^{2}\left(
0\right) \
\end{equation}

Also, the energy levels of the semi-infinite well, corresponding to the
limit $\,U_{0}\rightarrow \infty $ in (30), are given by:

\begin{equation}
E_{n}^{\left( s\infty \right) }=\frac{2\hbar ^{2}X_{2n-1}\left( 0\right) ^{2}%
}{ma^{2}}=\frac{2\hbar ^{2}\pi ^{2}}{ma^{2}}\left( n-\frac{1}{2}\right)
^{2}\
\end{equation}

Numerical studies show that the series expansion (53) can be used only in a
small interval near the origin; its value in another point must be obtained
constructing the series expansion near that point. This is why, for
practical calculations, it is convenient to use various approximations of
the exact formula.

Restricting the series (53) at its first 3 terms, we get Barker's
approximation \cite{[Barker1991]}:

\begin{equation}
X_{n}^{\left( B\right) }\left( p\right) =\frac{n\pi }{2}\left(
1-p+p^{2}-\left( 1+\frac{1}{6}\left( \frac{n\pi }{2}\right) ^{2}p^{3}\right)
\right) \
\end{equation}

Barker's approximation has been used in several applications of the square
well, for instance in the study of revivals and super-revivals. As given by
its author, the approximation formula is:

\begin{equation*}
\alpha _{n}\left( p\right) =\frac{1}{1+p}\frac{n\pi }{2}-\frac{p^{3}}{%
6\left( 1+p\right) ^{6}}\left( \frac{n\pi }{2}\right) ^{3}=
\end{equation*}

\begin{equation}
=\allowbreak \frac{1}{2}\pi n-\frac{1}{2}\pi np+\frac{1}{2}\pi np^{2}-\frac{1%
}{2}\pi n\left( 1+\frac{1}{24}\pi ^{2}n^{2}\right) p^{3}+...\
\end{equation}%
With our notations,

\begin{equation}
\alpha _{n}\left( p\right) =X_{n}^{\left( B\right) }\left( p\right) \
\end{equation}

As Barker's approximation is a consequence of the cubic approximation of the
$\sin $ function, its precision goes up to the third order terms, so the
terms not explicitly written in the r.h.s. of (62) must be omitted. In other
words, the correct form of the "Barker's approximation" is the polynomial
form given in (61), not the fractional form given in the original paper. We
can check this property verifying that

\begin{equation*}
\frac{\sin \left( \zeta _{1}^{\left( B\right) }\left( p\right) \right) }{%
\zeta _{1}^{\left( B\right) }\left( p\right) }=p+\mathcal{O}\left(
p^{4}\right) ,\ \ \frac{\sin \left( \zeta _{2}^{\left( B\right) }\left(
p\right) \right) }{\zeta _{2}^{\left( B\right) }\left( p\right) }=-p+%
\mathcal{O}\left( p^{4}\right)
\end{equation*}%
and so on. Such a property has any polynomial approximation of the exact
solution. For instance, putting:

\begin{equation*}
\zeta _{n}^{\left( N\right) }\left( p\right) =X_{2n}^{\left( N\right)
}\left( p\right) =n\pi \sum_{m=0}^{N}q_{m}\left( n\pi \right) p^{m}
\end{equation*}%
we get:

\begin{equation*}
\frac{\sin \left( \zeta _{n}^{\left( N\right) }\left( p\right) \right) }{%
\zeta _{n}^{\left( N\right) }\left( p\right) }=\left( -1\right) ^{n+1}p+%
\mathcal{O}\left( p^{N+1}\right)
\end{equation*}

Graphical and numerical considerations show that Barker's approximation
cannot be used for $\xi _{1}$ and $\zeta _{1},$ but is useful for any other
case, being more precise for smaller values of $p.$ The parabolic
approximation can be used, qualitatively, for $\zeta _{1},$ but is less
precise than Barker's one, in all other cases. The cubic approximation gives
excellent results in all cases, excepting $\xi _{1},$ where de Alcantara
Bonfim - Griffith solution should be used. Polynomial approximations of the
exact solutions, higher than Barker's, are useful only for small arguments,
so Barker's approximation can be considered as the most convenient
polynomial approximation. For the illustration of these conclusions, we give
here the plots for the $\zeta _{1}$ and $\zeta _{2}$ functions.

In Fig. 2, we can see how accurately the solutions $\zeta _{1}^{\left(
B\right) },\ \zeta _{1}^{\left( sp\right) }$ and $\zeta _{1}^{\left(
c\right) }$ satisfy the equation (23). The functions

\begin{equation*}
\frac{\sin \left( \zeta _{1}^{\left( B\right) }\left( p\right) \right) }{%
\zeta _{1}^{\left( B\right) }\left( p\right) },\ \frac{\sin \left( \zeta
_{1}^{\left( sp\right) }\left( p\right) \right) }{\zeta _{1}^{\left(
sp\right) }\left( p\right) },\ \frac{\sin \left( \zeta _{1}^{\left( c\right)
}\left( p\right) \right) }{\zeta _{1}^{\left( c\right) }\left( p\right) }
\end{equation*}%
are plotted with solid black, solid green and dotted black lines,
respectively; the function $f_{1}\left( p\right) =p$ - with red. Barker's
approximation is unusable, the parabolic one - quite poor, and the cubic one
- very good. In Fig. 3, we check, similarly, how accurately the solutions $%
\zeta _{2}^{\left( B\right) },\ \zeta _{2}^{\left( sp\right) }$ and $\zeta
_{2}^{\left( c\right) }$ satisfy the equation (24); the conventions are
similar, excepting the fact that instead of the function $f_{1},$ the
function $f_{2}\left( p\right) =-p$ is plotted. Now, the Barker
approximation is reasonably correct, especially for small values of the
argument, the parabolic one is poorer than Barker's, and the cubic one is
very accurate. These conclusions remain valid for the higher orders of both $%
\zeta $ and $\xi $ functions, with the remark that, in general, Barker's
approximations becomes quite poor for large values of the argument.

\begin{figure}[tbph]
\begin{center}
\epsfbox{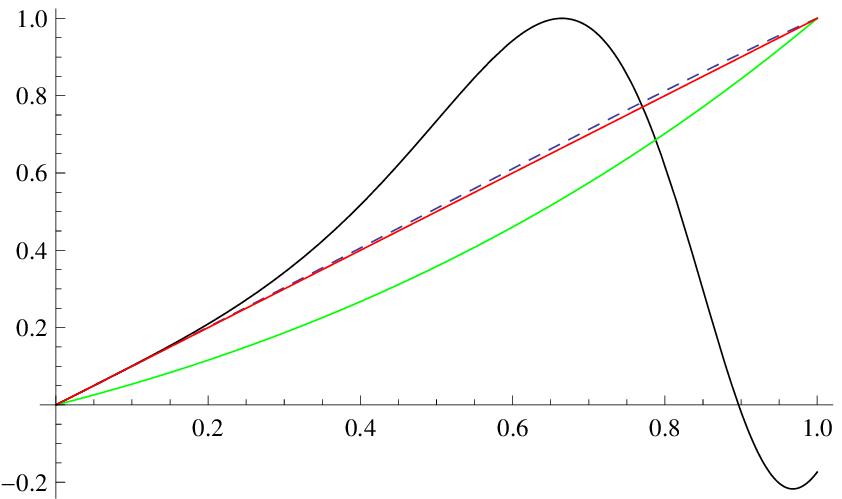}
\end{center}
\caption{The functions $\frac{\sin \left( \zeta _{1}^{\left(
B\right) }\left( p\right) \right) }{\zeta _{1}^{\left( B\right) }\left(
p\right) }$ (solid, black)$,\ \frac{\sin \left( \zeta _{1}^{\left( sp\right)
}\left( p\right) \right) }{\zeta _{1}^{\left( sp\right) }\left( p\right) }$
(solid, green)$,\ \frac{\sin \left( \zeta _{1}^{\left( c\right) }\left(
p\right) \right) }{\zeta _{1}^{\left( c\right) }\left( p\right) }$ (dotted,
black) and $f_{1}\left( p\right) =p$ (solid, red).}
\label{Fig2}
\end{figure}

\begin{figure}[tbph]
\begin{center}
\epsfbox{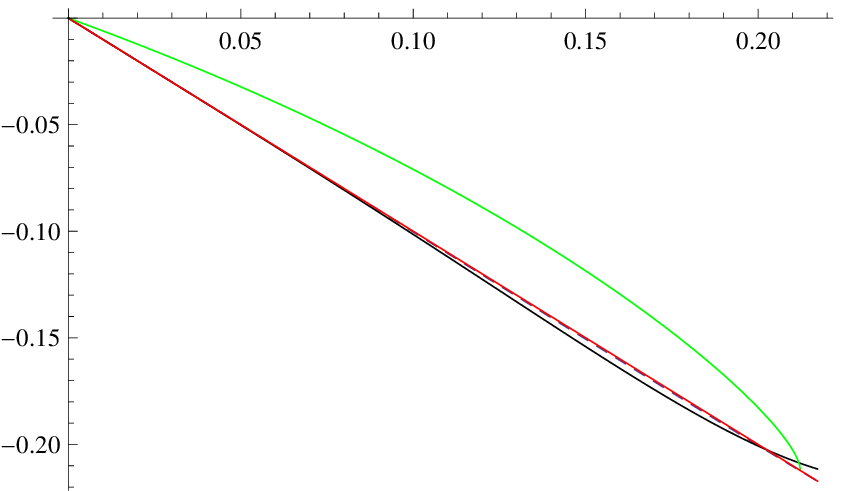}
\end{center}
\caption{The functions $\frac{\sin \left( \zeta _{2}^{\left(
B\right) }\left( p\right) \right) }{\zeta _{2}^{\left( B\right) }\left(
p\right) }$ (solid, black)$,\ \frac{\sin \left( \zeta _{2}^{\left( sp\right)
}\left( p\right) \right) }{\zeta _{2}^{\left( sp\right) }\left( p\right) }$
(solid, green)$,\ \frac{\sin \left( \zeta _{2}^{\left( c\right) }\left(
p\right) \right) }{\zeta _{2}^{\left( c\right) }\left( p\right) }$ (dotted,
black) and $f_{2}\left( p\right) =-p$ (solid, red).}
\label{Fig3}
\end{figure}

Maybe because the rectangular well is, from the point of view of advanced
physics, an elementary problem (even if, from the mathematical point of
view, the eigenvalue equations for the energy are not at all trivial), it is
sometimes underestimated, and several results obtained by various authors
remained unnoticed, and re-obtained in other forms; so, the domain has many
repetitions and overlappings. For instance, in \cite{[Pogosov2005]}, the
authors use the same method as that just exposed here in order to find a
series expansion for the solutions of the eigenvalue equations, but the
expansion used by them, equivalent to (65), is limited to second order
terms. In fact, the precision of their research, which is focused on the
physical properties of a mesoscopic film of thickness $L_{z}$, does not
request a better precision. However, this precision is inferior to the
precision of the Barker approximation which, as we could see, goes to third
order terms. Apparently, it seems that in \cite{[Pogosov2005]} Barker's
results are ignored, and re-obtain the series using a different method.

Also, in spite of the fact that the equivalence of the Sturm - Liouville
problems for electromagnetic waves and wave functions is well known since
the early years of the quantum mechanics, the results obtained in the square
well problem seem to remain unused by the researchers studying light
propagation in waveguides or in other simple geometries. Reciprocally, the
very detailed analysis of the eigenvalue equations, very similar to that of
a square well, used in the propagation of electromagnetic waves \cite%
{[Marcuse]}, where a number of approximate solutions have been proposed (see
for instance the references [90, 92] in \cite{[Marcuse]}), seem to have
remained unnoticed by people working in quantum mechanics.

\bigskip

\section{Applications: ultrathin metallic films, revivals and
electromagnetism}

\subsection{Ultra-thin metallic films}

Even if the infinite square well is a very crude model, it can explain
satisfactorily many aspects of the physics of metallic thin films. This is
not very surprising, as the energy levels of a deep square well are
acceptably approximated, for large quantum numbers, by the levels of an
infinite well. The approximation fails for the first ones $\left( n\lesssim
5\right) ,$ but for a film with $n\gtrsim $ 25 monolayers, the effect is not
dramatic. The infinite well model has been improved, taking into account, in
an approximative manner, the fact that the electron moving in the well can
penetrate its walls; this is the so-called phase accumulation model (PAM).
The model is exposed in a number of papers, \cite{[Smith1985]}, \cite%
{[Milun2002]},\ \cite{[Qiu2002]}; concerning PAM for a metal - semiconductor
interface, see \cite{[Tang2011]}. However, even PAM gives very inaccurate
results for very thin films. The fact that both the infinite well model and
PAM are inappropriate for shallow wells (for instance, in an 1 ML Ag film,
only a single QW state exists \cite{[Milun2002]}) is discussed by several
researchers. For instance, in \cite{[Wu2008]}, the authors warn about the
fact that we have to be very cautious when we apply the infinite well model
to very thin films; in fact, they find that the stability of a thin film of
less than 5 monolayers depends sensitively on the height of interface energy
barriers, (see \cite{[Wu2008]}, p.6). In the photoemission studies of QW
states in thin films, reported in \cite{[ChiangSSR2000]} and \cite%
{[Chiang2000PRB]}, the data for the first 3 monolayers are excluded from the
fit, in a PAM analysis. The angle - resolved photoemission is indeed an
important tool for the study of electron states in QWs. In fact, there is a
close analogy between a standing electromagnetic wave propagating between
two reflecting surfaces and an electron moving in a square potential well
\cite{[Cohen-Tannoudji]}. The first case corresponds to the Fabry - Perot
interferometer, which shows a set of peaks in optical transmission; the
second, to an electron confined in QW, formed in an ultra-thin metallic
film. Electron confinement within a film results in discrete QW states,
observable by angle-resolved photoemission. For a such film of Ag on Fe,
angle-resolved photoemission spectra are well explained by PAM for number of
ML >2, but fails for 1 or 2 monolayers. \cite{[Paggel1999]}.

So, the exact treatment of the finite well, or analytic approximations of
the exact solutions are necessary for a correct description of the shallow
wells (presuming that the model itself is still usable in all ultra-thin
films). Our results can be directly used in the approach of \ Wu and Zhang
\cite{[Wu2008]} who apply the finite well model to the study of very thin
films, but use numerical instead of analytical solutions. For freestanding
films, where the electrons move in a symmetric square well of barrier height

\begin{equation}
V_{0}=W_{m}+E_{F}\
\end{equation}%
the theory works without any adjustable parameter. The freestanding films
are of great theoretical interest, as they can serve as reference points to
see how substrates affect film properties.

For a film on a semiconductor substrate, the film is modeled as electrons
residing in an asymmetric well, so:

\begin{equation}
V\left( x<\left\vert a\right\vert /2\right) =V_{1}=W_{m1}+E_{F},\ \ \ \
V\left( x>\left\vert a\right\vert /2\right) =V_{2}=W_{m2}+E_{F}\
\end{equation}%
As $W_{m1},\ W_{m1}\ $are comparable, $\left\vert V_{1}-V_{2}\right\vert \ll
V_{1},\ V_{2},$ and the eigenenergy can be obtained using the perturbation
theory, considering the symmetric well as exactly solvable problem. The same
remarks applies for an electron in a corrugated box, for instance if the
corrugated periodic potential is (\cite{[Wu2008]}, eq. (29)):

\begin{equation}
V_{c}\left( z\right) =v\cos \frac{2\pi z}{d_{0}}\
\end{equation}%
with $\nu \ll V_{0}.$

For solid-solid and solid-vacuum interfaces, the confinement potential is
generally finite and rounded. Examples of such potentials, for multilayered
nanostructures, obtained using the dynamical mean-field theory approach, can
be found in \cite{[Freericks]}, Sect. 3.6. If this more realistic form of
the potential has to be taken into account, its eigenstates can be obtained
by simple perturbation calculation, from the eigenstates of the rectangular
well.

It is well known that the Sturm - Liouville problems in optics and quantum
mechanics can be identical \cite{[Black1985]}, \cite{[Barsan-Electromg]}, so
the quantized values of the wave vector of the quantum mechanical problem
correspond perfectly to the normal modes of transversal electric field in
resonant cavities, wave guides or heterostructure lasers, so our solutions
of the energy eigenvalues can be applied also in optics and electromagnetism.

\subsection{Revivals}

As it is well known, wave packets in infinite square wells produce revivals,
and in finite square wells - both revivals and super-revivals. Analyzing
numerically the superrevivals in finite square wells, \cite%
{[Venugopalan1999]} notices that the Barker's formula:

\begin{equation}
E_{n}\simeq \frac{P^{2}}{\left( P+1\right) ^{2}}\frac{\pi ^{2}\hbar ^{2}}{%
2mL^{2}}n^{2}\
\end{equation}%
(actually, the first term of the Barker's formula, obtained in fact by
Garrett \cite{[Garrett]}) is less precise for shallow wells, than for deep
wells. This property clearly results from our analysis of exact and
approximate solutions of the square well problem. The same approximation is
used by \cite{[Aronstein1997]} However, no analytic result can be obtained
in this way, as the $n-$dependence of the eigenenergy is qualitatively
different in the two cases (for the finite square well, the simple $n^{2}$
dependence disappears), and this difference is smeared out in the Barker
approximation.

The solutions of the transcendental equations allows us to find the analytic
expressions of the revivals and superrevivals times for the finite square
well, according to the Eqs. (10), (26-28) of \cite{[AronsteinAJP2000]}.

\subsection{Electromagnetism}

The electromagnetic wave propagation in the symmetric three-layer slab
waveguide is an example of phenomena described by a Sturm - Liouville system
essentially identical to that of the quantum square well. Due to the large
number of quantities to be defined in order to follow in detail the
similarity of the two phenomena, we shall omit a detailed presentation,
mentioning only that the equations giving the propagation constant (the wave
vector) of the even and odd transversal electric modes, and of the even and
odd transversal magnetic modes, respectively (2.4-45), (2.4-54); (2.4-60),
(2.4-66) in \cite{[Casey1978]}\ correspond exactly to the eigenvalue
equations for the energy of even and odd bound states in a quantum square
well, respectively Eqs. (25.4e) and (25.4o) of \cite{[Fluegge1971]}.

\bigskip

\section{Conclusions}

This paper represents an illustration of the situation when solutions of
simple problems of quantum mechanics can be used for description of
interesting phenomena in nanophysics. Specifically, we have referred to
exact solutions of the eigenvalue equations for the energy of bound states
of a particle in a rectangular well. Even if the physics of this problem is
elementary, its mathematics is not trivial, and consists in solving a
transcendental equation. Both exact and approximate solutions are obtained,
with various degrees of accuracy. The results are interesting for the
physics of quantum wells, mainly for the important case of ultra-thin
metallic films. Besides the case of electrons in rectangular quantum wells,
the case of rugose bottom wells can be treated, with an elementary
perturbation theory. Also, the theory described here has applications in the
domain of heterojunction lasers and in the electromagnetic wave propagation
in three slab dielectric waveguides.

\bigskip

\appendix

\section{Appendix}

The function $\xi _{1}$ cannot be obtained using the polynomial
approximations described in Section 3. Very accurate analytical
approximations for $\xi _{1}$ (with an error of about $10^{-3}$) have been
proposed by de Alcantara Bonfim and Griffith \cite{[Alcantara2006]}, using
an algebraic approximations for the $\cos $ function.

The cubic approximation for the following 4 $\xi _{n}$ functions is:

\begin{equation}
\xi _{2}^{\left( c\right) }\left( p\right) =4.2409+2.8851\sin \left( \frac{1%
}{3}\arcsin \left( 0.4728-4.3767p\right) \right) \   \tag{A1}
\end{equation}

\begin{equation}
\xi _{3}^{\left( c\right) }\left( p\right) =7.6132+2.9840\sin \left( \frac{1%
}{3}\arcsin \left( 0.2300-7.6906p\right) \right) \   \tag{A2}
\end{equation}

\begin{equation}
\xi _{4}^{\left( c\right) }\left( p\right) =10.8580+3.0803\sin \left( \frac{1%
}{3}\arcsin \left( 0.1336-10.6235p\right) \right) \   \tag{A3}
\end{equation}

\begin{equation}
\xi _{5}^{\left( c\right) }\left( p\right) =14.0607+3.1484\sin \left( \frac{1%
}{3}\arcsin \left( 0.0729-13.439p\right) \right) \   \tag{A4}
\end{equation}

The cubic approximation for $\zeta _{1}$ is given in Section 3. The next
four $\zeta _{n}$ functions are:

\begin{equation}
\zeta _{2}^{\left( c\right) }\left( p\right) =5.9562+2.9256\sin \left( \frac{%
1}{3}\arcsin \left( 0.3297-6.121p\right) \right)  \tag{A5}
\end{equation}

\begin{equation}
\zeta _{3}^{\left( c\right) }\left( p\right) =9.24337+3.03623\sin \left(
\frac{1}{3}\arcsin \left( 0.178\,39-9.\,\allowbreak 179\,3p\right) \right)
\tag{A6}
\end{equation}

\begin{equation}
\zeta _{4}^{\left( c\right) }\left( p\right) =12.4627+3.1173\sin \left(
\frac{1}{3}\arcsin \left( 0.0996-12.0402p\right) \right) \   \tag{A7}
\end{equation}

\begin{equation}
\zeta _{5}^{\left( c\right) }\left( p\right) =15.6911+3.2908\sin \left(
\frac{1}{3}\arcsin \left( 0.01538-14.3186p\right) \right)  \tag{A8}
\end{equation}

\textbf{The coefficients }$q_{7},...q_{16},$\textbf{\ of the exact series
expansion (53)}

\bigskip

\begin{equation}
q_{7}\left( b\right) =-\left( 1+\frac{5\cdot 7}{2\cdot 3}b^{2}+\frac{7\cdot
37}{2^{3}\cdot 3\cdot 5}b^{4}+\frac{5}{2^{7}\cdot 7}b^{6}\right) \   \tag{A9}
\end{equation}

\begin{equation}
q_{8}\left( b\right) =\left( 1+\frac{2^{2}\cdot 7}{3}b^{2}+\frac{2\cdot 7^{2}%
}{3\cdot 5}b^{4}+\frac{2^{4}}{5\cdot 7}b^{6}\right)  \tag{A10}
\end{equation}

\begin{equation}
q_{9}\left( b\right) =-\left( 1+2\cdot 7b^{2}+\frac{7\cdot 47}{2^{2}\cdot 5}%
b^{4}+\frac{3229}{2^{2}\cdot 3^{2}\cdot 5\cdot 7}b^{6}+\frac{5\cdot 7}{%
2^{7}\cdot 3^{2}}b^{8}\right) \   \tag{A11}
\end{equation}

\begin{equation}
q_{10}\left( b\right) =\left( 1+2^{2}\cdot 5b^{2}+\frac{2\cdot 7\cdot 13}{5}%
b^{4}+\frac{2^{4}\cdot 41}{3^{2}\cdot 7}b^{6}+\frac{2^{7}}{3^{2}\cdot 5\cdot
7}b^{8}\right) \   \tag{A12}
\end{equation}

\begin{equation}
q_{11}\left( b\right) =-\left( 1+\frac{5\cdot 11}{2}b^{2}+\frac{7\cdot
11\cdot 19}{2^{2}\cdot 5}b^{4}+\frac{11\cdot 1571}{2^{3}\cdot 3^{2}\cdot 7}%
b^{6}+\frac{11\cdot 59\cdot 181}{2^{7}\cdot 3^{2}\cdot 5\cdot 7}b^{8}+\frac{%
3^{2}\cdot 7}{2^{8}\cdot 11}b^{10}\right) \ \   \tag{A13}
\end{equation}

\begin{equation}
q_{12}\left( b\right) =\left( 1+\frac{2\cdot 5\cdot 11}{3}b^{2}+\frac{2\cdot
11\cdot 31}{5}b^{4}+\frac{2^{2}\cdot 11\cdot 139}{3^{2}\cdot 7}b^{6}+\frac{%
2^{3}\cdot 11\cdot 479}{3^{4}\cdot 5\cdot 7}b^{8}+\frac{2^{8}}{3^{2}\cdot
7\cdot 11}b^{10}\right) \ \   \tag{A14}
\end{equation}

\begin{equation*}
q_{13}\left( b\right) =-\left( 1+\frac{11\cdot 13}{3}b^{2}+\frac{11\cdot
13\cdot 67}{2^{5}\cdot 5}b^{4}+\frac{11\cdot 13\cdot 17\cdot 127}{2^{2}\cdot
3^{2}\cdot 5\cdot 7}b^{6}+\frac{11\cdot 13\cdot 23\cdot 6679}{2^{7}\cdot
3^{4}\cdot 5\cdot 7}b^{8}+\right.
\end{equation*}

\begin{equation}
\left. +\frac{13\cdot 211\cdot 2609}{2^{7}\cdot 3^{2}\cdot 5^{2}\cdot 7\cdot
11}b^{10}+\frac{3\cdot 7\cdot 11}{2^{10}\cdot 13}b^{12}\right) \ \ \
\tag{A15}
\end{equation}

\begin{equation*}
q_{14}\left( b\right) =\left( 1+\frac{2\cdot 7\cdot 13}{3}b^{2}+\frac{2\cdot
7\cdot 11\cdot 13}{5}b^{4}+\frac{2^{2}\cdot 11\cdot 13\cdot 311}{3^{2}\cdot
5\cdot 7}b^{6}+\frac{2^{4}\cdot 11\cdot 13\cdot 37}{3^{4}\cdot 5}%
b^{8}+\right.
\end{equation*}

\begin{equation}
\left. \frac{2^{6}\cdot 13\cdot 59}{3^{2}\cdot 5^{2}\cdot 11}b^{10}+\frac{%
2^{10}}{3\cdot 7\cdot 11\cdot 13}b^{12}\right) \   \tag{A16}
\end{equation}

\begin{equation*}
q_{15}\left( b\right) =-\left( 1+\frac{5\cdot 7\cdot 13}{2\cdot 3}b^{2}+%
\frac{7^{2}\cdot 11^{2}\cdot 13}{2^{3}\cdot 3\cdot 5}b^{4}+\frac{11\cdot
13\cdot 8521}{2^{4}\cdot 3^{2}\cdot 7}b^{6}+\frac{11\cdot 13\cdot 79\cdot
2917}{2^{7}\cdot 3^{4}\cdot 5}b^{8}+\right.
\end{equation*}

\begin{equation}
\left. \frac{7\cdot 13\cdot 1206053}{2^{8}\cdot 3^{4}\cdot 5\cdot 11}b^{10}+%
\frac{17911\cdot 135721}{2^{10}\cdot 3^{3}\cdot 5^{2}\cdot 7\cdot 11\cdot 13}%
b^{12}+\frac{11\cdot 13}{2^{11}\cdot 5}b^{14}\right) \   \tag{A17}
\end{equation}

\begin{equation*}
q_{16}\left( b\right) =\left( 1+\frac{2^{3}\cdot 5\cdot 7}{3}b^{2}+\frac{%
2^{2}\cdot 7\cdot 13\cdot 41}{3\cdot 5}b^{4}+\frac{2^{4}\cdot 11\cdot
13\cdot 67}{3^{2}\cdot 5\cdot 7}b^{6}+\frac{2\cdot 11\cdot 13\cdot 2473}{%
3^{4}\cdot 5}b^{8}+\right.
\end{equation*}

\begin{equation}
\left. \frac{2^{5}\cdot 13\cdot 4201}{3^{4}\cdot 5\cdot 11}b^{10}+\frac{%
2^{6}\cdot 266681}{3^{3}\cdot 5^{2}\cdot 7\cdot 11\cdot 13}b^{12}+\frac{%
2^{11}}{3^{2}\cdot 5\cdot 11\cdot 13}b^{14}\right) \   \tag{A18}
\end{equation}

\begin{acknowledgement}
The financial support of the ANCS project PN 09 37 01 06, of the JINR Dubna
- IFIN-HH Magurele-Bucharest project no. 01-3-1072-2009/2013 and of the CEI
Cooperation Fund Project No. 1202.161-13 are kindly acknowledged.
\end{acknowledgement}


\bigskip

\begin{thebibliography}{99}
\bibitem{[Mitin2010]} V. V. Mitin, D. I. Sementsov, N. Z. Vagidov: Quantum
Mechanics for Nanostructures, Cambridge University Press (2010)

\bibitem{[VB-RD]} V. Barsan, R. Dragomir, Optoel.Adv.Mat - Rapid Commun.
\textbf{6}, 917 (2012)

\bibitem{[Harrison]} W.A. Harrison: Applied Quantum Mechanics, World
Scientific (2000)

\bibitem{[Van Wyk2011]} S. Van Wyk: Computer Solutions in Physics, World
Scientific (2011)

\bibitem{[Fluegge1971]} S. Fluegge: Practical Quantum Mechanics,
Springer-Verlag Berlin Heidelberg New York (1971)

\bibitem{[Ando1982]} T. Ando, A. B. Fowler, F. Stern, Rev.Mod.Phys. \textbf{%
54}, 437 (1982)

\bibitem{[Kolbas1984]} R. M. Kolbas, N. Holonyak, Jr.(1984), Amer.J.Phys.
\textbf{52}, 431 (1984)

\bibitem{[Wu2008]} B. Wu, Z. Zhang, Phys.Rev.\textbf{B77}, 035410 (2008)

\bibitem{[Czosche2003]} P. Czosche et al., Phys.Rev.Lett. \textbf{91},
226801 (2003)

\bibitem{[Kim2010]} J. Kim et al., Proc.Natl.Ac.Sci. \textbf{107}, 12761
(2010)

\bibitem{[Sandomirskii1967]} V.B. Sandomirskii, JETF \textbf{52}, 158
(1967); Sov.Phys. JETP \textbf{25}, 101 (1967)

\bibitem{[Schulte1976]} F.K. Schulte, Surf.Sci. \textbf{55}, 427 (1976)

\bibitem{[Qiu2002]} Z. Q. Qiu, N. V. Smith: J.Phys.:Cond.Matter \textbf{14},
R169 (2002)

\bibitem{[Atkinson2008]} W. A. Atkinson, A. J. Slavin, Amer.J.Phys. 74 43
(2008)

\bibitem{[Barker1991]} B. I. Barker, G. H. Rayborn, J. W. Ioup, G. E. Ioup,
Am.J.Phys. \textbf{59}, 1038 (1991)

\bibitem{[AronsteinAJP2000]} D. L. Aronstein, C. R. Stroud, Amer.J.Phys.
\textbf{68}, 943 (2000)

\bibitem{[Alcantara2006]} O. F. de Alcantara Bonfim, D. J. Griffiths,
Amer.J.Phys. \textbf{74}, 43 (2006)

\bibitem{[VB-RRP]} V. Barsan, Rom.Rep.Phys. \textbf{64}, 685 (2012)

\bibitem{[Pogosov2005]} V. V. Pogosov, V. P. Kurbatsky, E. V. Vasyutin,
Phys.Rev.B195410 (2005)

\bibitem{[Marcuse]} D. Marcuse, Light transmission optics, Van Nostrand, New
York (1982)

\bibitem{[Smith1985]} N. V. Smith, PhysRev \textbf{B32}, 3549 (1985)

\bibitem{[Milun2002]} M. Milun, P. Pervan, D.P. Woodruff, Rep.Prog.Phys.
\textbf{65}, 99 (2002)

\bibitem{[Tang2011]} S.-J. Tang et al. Phys.Rev.Lett. \textbf{107}, 066802
(2011)

\bibitem{[ChiangSSR2000]} T.-C. Chiang, Photoemission studies of QW states
in thin films, Surf.Sci.Rep.\textbf{39}, 181 (2000)] and [Chiang2000PRB]
PhysRev \textbf{B61} 1804 (2000)

\bibitem{[Chiang2000PRB]} T.-C. Chiang, Phys.Rev. \textbf{B61,} 1804 (2000)

\bibitem{[Cohen-Tannoudji]} C. Cohen-Tannoundji, B. Diu, F. Laloe: Quantum
Mechanics, Wiley, New York (1977)

\bibitem{[Paggel1999]} J. J. Paggel, T. Miller, T.-C. Chiang, Science
\textbf{283}, 1709 (1999)

\bibitem{[Freericks]} J. K. Freericks: Transport in multilayered
nanostructures, Imperial College Press, 2006

\bibitem{[Black1985]} R. J. Black, A. Ankiewicz, Am.J.Phys.\textbf{53}, 554
(1985)

\bibitem{[Barsan-Electromg]} V. Barsan: Waveguides, resonant cavities,
optical fibers and their quantum counterparts, in: V. Barsan, R. P. Lungu
(Eds.): Trends in electromagnetism, InTech, 2011

\bibitem{[Venugopalan1999]} A. Venugopalan, A. G. Agarwal, Phys.Rev.\textbf{%
A59}, 1413 (1999)

\bibitem{[Garrett]} S. Garrett, Amer.J.Phys. \textbf{47}, 195 (1979)

\bibitem{[Aronstein1997]} D. L. Aronstein, C. R. Stroud, Phys.Rev.\textbf{A55%
}, 4526 (1997)

\bibitem{[Casey1978]} H. C. Casey, Jr., M. B. Panish: Heterostructure
lasers, Academic Press, New York, San Francisco, London (1978)
\end{thebibliography}
\end{document}